\documentclass[prd,preprint,aps,tightenlines,showpacs,superscriptaddress]{revtex4-1}
\usepackage[dvips]{graphicx}
\usepackage{dcolumn}
\usepackage{epsfig}
\usepackage{amssymb}

\newcommand{\ket}[1]{|{#1}\rangle}
\newcommand{\bra}[1]{\langle{#1}|}
\newcommand{\inp}[2]{\langle{#1}|{#2}\rangle}

\begin{document}

\title{Neutrino-induced forward meson-production reactions in nucleon resonance region}

\author{H. Kamano}
\affiliation{Research Center for Nuclear Physics, Osaka University, Ibaraki, Osaka 567-0047, Japan}
\author{S. X. Nakamura}
\affiliation{Yukawa Institute for Theoretical Physics, Kyoto University, Kyoto 606-8542, Japan}
\author{T.-S. H. Lee}
\affiliation{Physics Division, Argonne National Laboratory, Argonne, Illinois 60439, USA}
\author{T. Sato}
\affiliation{Department of Physics, Osaka University, Toyonaka, Osaka 560-0043, Japan}
\affiliation{J-PARC Branch, KEK Theory Center, Institute of Particle and Nuclear Studies, High Energy Accelerator Research Organization (KEK), Tokai, Ibaraki 319-1106, Japan}

\begin{abstract}
As a first step toward developing a reaction model that enables a comprehensive 
description of neutrino-nucleon reactions in the nucleon resonance region, 
we have applied for the first time a dynamical coupled-channels model,
which successfully describes $\pi N, \gamma N \to \pi N, \eta N, \pi \pi N, K\Lambda, K\Sigma$ 
reactions up to $W = 2 $ GeV, 
to predict the neutrino-induced meson-production reactions with $\Delta S = 0$ 
at the forward angle limit.
This has been achieved by relating the divergence of the axial-current matrix elements at $Q^2 = 0$ 
to the $\pi N \to X$ reaction amplitudes through the PCAC hypothesis.
We present the contributions from each of the 
$\pi N, \eta N, \pi \pi N, K\Lambda, K\Sigma$ channels to the
$F_2$ structure function at $Q^2 \to 0$ limit up to $W = 2$ GeV. 
\end{abstract}

\maketitle

\section{Introduction}
\label{sec:int}

Recent breakthrough measurements of non-zero neutrino mixing angle $\theta_{13}$
from Daya Bay and RENO experiments~\cite{daya-bay,reno}, which are consistent with
the data from T2K, MINOS and Double Chooz experiments~\cite{t2k,minos,dc},
indicated a possibility of the CP violation in the lepton sector.
Now the main issue of the neutrino physics is shifting to CP phase,
mass hierarchy as well as further precise determination of $\theta_{13}$.
For making a progress towards this direction by analyzing data from
the next-generation long-baseline and atmospheric experiments,
neutrino-nucleon and neutrino-nucleus scattering need to be
understood within 10\% or better accuracy, for the relevant neutrino energy
region from sub GeV to a few GeV, and $0\le Q^2\le 4$ (GeV/c)$^2$ 
[see Eq.~(\ref{eq:Q2}) for the definition of  $Q^2$].
This energy region covers neutrino-nucleus interactions of different
characteristics, namely, the quasi-elastic (QE), resonant (RES), and
deep-inelastic scatterings (DIS).
Thus a combination of different expertise is necessary to tackle the problem. 
This motivates theorists and experimentalists to get together to
organize a new collaboration, e.g., see Ref.~\cite{jparc-kek}.

Here we are concerned with the RES region
which covers the $\Delta$ peak and, through the second and third
resonance regions, up to the region overlapping with the DIS region.
Previous models for the weak single pion production off the nucleon in
the RES region, some of them are for the
$\Delta$-region only, can be categorized into three kinds of approaches.
Models of the first kind of approaches consist of a coherent sum of resonance
contributions~\cite{rs1,rs1-2,lalakulich1,lalakulich1-2}.
The second one additionally has non-resonant mechanisms of the tree
level~\cite{hernandez1,hernandez1-2,lalakulich2}.
The third one considers the rescattering also so that the $\pi N$ unitarity is
maintained, and such a model for the $\Delta$-region
was developed by two of the present authors~\cite{sl3,msl05}.
These models for the elementary processes have been used as basic ingredients
to construct neutrino-nucleus reaction models.
Although the previous models mentioned above consider only the single-pion production,
double-pion production is comparably important in the RES region.
Furthermore, $\eta$ and kaon productions also take place, and they can be a
background for proton-decay experiments~\cite{proton-decay1,proton-decay2}.
Some models for the weak kaon productions 
through the strangeness conserving ($\Delta S = 0$) reactions~\cite{adera}
and the strangeness changing ($\Delta S =\pm 1$) reactions~\cite{rafi12,rafi10},
belonging to the second kind of approaches discussed above,
have been developed so far.
In order to describe those meson production reactions,
the reaction model has to take into account the coupled-channels effects and
satisfy unitarity for the multichannel reactions.
However, such a model for the
neutrino-nucleon reactions has not been developed so far.

In this context, our recent development of a dynamical coupled-channels (DCC)
model is quite encouraging~\cite{aip11,knls12}.
Our DCC model is based on a comprehensive analysis of 
$\pi N,\gamma N\to \pi N, \eta N, K\Lambda, K\Sigma$ reactions in the
RES region, taking account of the coupled-channels unitarity including 
the $\pi\pi N$ channel.
An extension of the DCC model to the neutrino reaction is fairly
straightforward.
Although we need to construct a dynamical axial-current model 
for a full development, we can actually calculate the neutrino-induced
forward ($Q^2=0$) meson production cross sections, characterized by the
structure function $F_2$,
from the cross sections for $\pi N\to X$ ($X=\pi N, \eta N, KY ...$) 
by invoking the PCAC hypothesis.
Thus, in this report, we attempt to make a first step of extending the
DCC model to the weak sector, by calculating $F_2(Q^2=0)$ for 
$\Delta S = 0$ $\nu N\to l X$ ($l$: lepton) with
the PCAC hypothesis, thereby setting a starting point for a future full development.
We also remark that our estimate of the magnitudes of 
$\nu N\to l KY$ and $\nu N\to l \eta N$ forward cross sections 
is, for the first time, based on a model that has been 
rather extensively tested by the data of $\pi N$ and $\gamma N$ reactions in the RES region.

The rest of this report is organized as follows:
In Sec.~\ref{sec:form}, we describe our procedure to calculate $F_2$ for the
forward neutrino-induced meson production reaction using the PCAC hypothesis.
We present numerical results in Sec.~\ref{sec:result}, followed by a
summary in Sec.~\ref{sec:summary}.

\section{Formalism}
\label{sec:form}

\subsection{Kinematics}

First we define kinematic variables needed for the following discussions.
We consider the inclusive $l (k) + N(p) \to l'(k') + X(p')$ reactions, where 
$(l,l') = (\nu_e, e^-), (\bar \nu_e, e^+)$ for the charged-current (CC) reactions,
while $(l,l') = (\nu_e, \nu_e), (\bar \nu_e, \bar \nu_e)$ for the neutral-current (NC) 
reactions.
We assume that leptons are massless throughout this paper.

In the laboratory frame, the four-momentum are defined to be
\begin{eqnarray}
k &=& (E, \vec k), 
\label{eq:k-lab}
\\
p &=& (m_N, 0, 0, 0),
\label{eq:p-lab}
\\
k' &=& (E', \vec k'),
\label{eq:k'-lab}
\end{eqnarray}
and $p' = k + p -k'$.
For massless leptons, $E=|\vec k|$ and $E'=|\vec k'|$.
The positive quantity $Q^2$ is then defined by 
\begin{equation}
Q^2 = -q^2 = 4 E E^\prime \sin^2\frac{\theta}{2} \,,
\label{eq:Q2}
\end{equation}
where $\theta$ is the scattering angle of $l'$ with respect to $l$, i.e.,
$\hat k \cdot \hat k' = \cos\theta$;
$q$ is the momentum transfer between $l$ and $l'$, $q= k'-k$. 
Each component of the four-momentum $q$ is denoted as $q=(\omega,\vec q)$ in the laboratory frame.
Hereafter we call this frame FM1.

For later use, we also define another frame, called FM2, in which $X$ is at rest.
In this frame, $q$ and $p$ are denoted as $ q=(\omega_c,\vec q_c)$ and 
$p = (E_N, -\vec q_c)$, respectively, where $E_N = \sqrt{m_N^2 + |\vec q_c|^2}$ 
and $m_N$ is the nucleon mass. Also, we set $\vec q_c =(0,0,|\vec q_c|)$ so that
$\vec q_c$ defines the $z$-direction of this frame.

\subsection{Cross section formula of inclusive neutrino reactions at forward angle limit}
\label{sec:form-1}

\begin{figure}[t]
\begin{center}
\includegraphics[width=0.5\textwidth,clip]{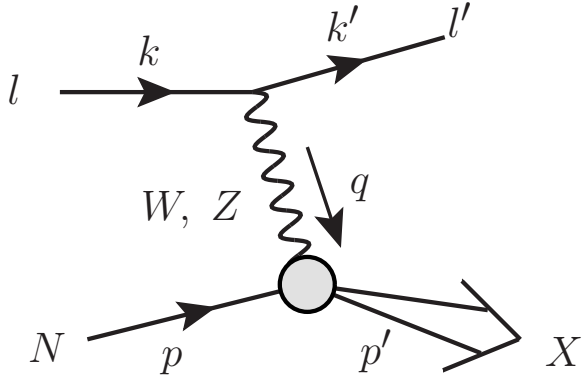}
\caption{
Schematic representation of the $l(k) + N(p) \to l'(k') + X(p')$ reactions
considered in this paper.
}
\label{fig:diag}
\end{center}
\end{figure}

By assuming that the inclusive $l (k) + N(p) \to l'(k') + X(p')$ reactions
take place via one-gauge-boson exchange processes as shown in Fig.~\ref{fig:diag},
the cross sections for the inclusive neutrino and anti-neutrino reactions are expressed as
\begin{equation}
\frac{d\sigma_\alpha}{dE^\prime d\Omega^\prime}
=
\frac{G^2_F C_\alpha}{2\pi^2}E^{\prime 2}
\left[ 
2W_{1,\alpha} \sin^2\frac{\theta}{2}
+W_{2,\alpha} \cos^2\frac{\theta}{2}
\pm W_{3,\alpha} \frac{E+E^\prime}{m_N}\sin^2\frac{\theta}{2}
\right] \,.
\label{eq:crs1}
\end{equation}
Here, the label $\alpha = \text{CC}\nu,\text{CC}\bar\nu,~\text{NC}\nu,~\text{NC}\bar\nu$ 
specifies the reactions;
$\Omega'$ is the solid angle of $l'$ in the laboratory frame;
$C_\alpha = |V_{ud}|^2$ for $\alpha = \text{CC}\nu,~\text{CC}\bar\nu$ and 
$C_\alpha = 1$ for $\alpha = \text{NC}\nu,~\text{NC}\bar\nu$;
the sign in front of $W_{3,\alpha}$ is taken to be $+$ ($-$) for $\nu$ ($\bar\nu$)
induced reactions.
The structure functions, $W_{i,\alpha}$ ($i=1,2,3$), are Lorentz-invariant functions of 
two independent variables. 
One usually chooses $Q^2$ and the invariant mass $W=\sqrt{s}=\sqrt{(p+q)^2}$
for the resonance region, but chooses Bjorken scaling variable $x=Q^2/(2p\cdot q)$ 
and $Q^2$ for the deeply inelastic region. 
In the forward limit, $\theta \to 0$, Eq.~(\ref{eq:crs1}) reduces to
\begin{equation}
\frac{d\sigma_\alpha}{dE^\prime d\Omega^\prime} (\theta \to 0)
=
\frac{G^2_F C_\alpha}{2\pi^2}E^{\prime 2} W_{2,\alpha}  \,.
\label{eq:crs2}
\end{equation}

The structure function $W_{2,\alpha}$ is expressed in terms of matrix elements of 
weak currents between the initial nucleon $N$ and the final $X$,
$\bra{X}J^\mu_\alpha\ket{N}$. 
[Throughout this paper, we use conventions of Bjorken and Drell~\cite{bd-book}, 
and any one-particle states
are normalized as $\inp{k}{k'} = \delta(\vec k - \vec k')$.]
The weak currents $J_{\alpha,\mu}$ are given by
\begin{equation}
J_{\alpha,\mu} =
\left\{
\begin{array}{cl}
(V^1_\mu + iV^2_\mu) - (A^1_\mu+iA^2_\mu)&(\text{for }\alpha =\text{CC}\nu), \\
(V^1_\mu - iV^2_\mu) - (A^1_\mu-iA^2_\mu)&(\text{for }\alpha =\text{CC}\bar\nu), \\
(1-2\sin^2\theta_W)V^3_\mu - 2\sin^2\theta_W V^{\text{IS}}_\mu- A^3_\mu& (\text{for }\alpha = \text{NC}\nu,~\text{NC}\bar\nu).
\end{array}
\right.
\end{equation}
Here, $V^i_\mu$ and $A^i_\mu$ are the vector and axial currents, respectively.
The superscript $i = \text{IS}$ ($i=1,2,3$)
denotes the isoscalar current ($i$-th component of the isovector current).
Also, $\theta_W$ is the Weinberg angle.
If one evaluates $\bra{X}J^\mu_\alpha\ket{N}$ in the frame FM2, then the structure functions 
are expressed as~\cite{msl05}
\begin{equation}
W_{2,\alpha} 
=
\frac{Q^2}{\vec q^2} \sum 
\left[ \frac{1}{2}\left( |\bra{X} J_\alpha^x \ket{N}|^2 + |\bra{X} J_\alpha^y \ket{N}|^2 \right)
+ \frac{Q^2}{\vec q_c^2} \left|\bra{X}
\left(J_\alpha^0 + \frac{\omega_c}{Q^2} q\cdot J_\alpha \right)
\ket{N}\right|^2 \right] \,,
\label{eq:w2-1}
\end{equation}
where we have introduced concise notation
\begin{equation}
\sum = \frac{1}{2}\sum_{N\text{-spin}} \sum_X (2\pi)^3 \delta^4(p + q - p')\frac{E_N}{m_N} \,,
\label{eq:sum}
\end{equation}
where $\sum_X$ means summing up all possible quantum numbers and integrating 
over momentum $\vec p'$ of all final state $X$, and the factor 
$1/2$ in Eq.~(\ref{eq:sum}) comes from taking average for the initial nucleon spin.

We now notice from Eq.~(\ref{eq:Q2}) that the $\theta \to 0$ limit leads to $Q^2 \to 0$, 
and thus the structure function
$W_{2,\alpha}$ for evaluating  the cross section [Eq.~(\ref{eq:crs2})] at $\theta \to 0$ reduces to
\begin{eqnarray}
 W_{2,\alpha}(Q^2 \to 0)  
= \frac{Q^2}{\vec q^2} \sum \frac{Q^2}{\vec{q}^2_c}
 |\bra{X} \frac{\omega_c}{Q^2} q\cdot J_\alpha 
\ket{N}|^2  .
\end{eqnarray}
Because of the vector current conservation $\bra{X} q\cdot V_{\alpha}\ket{N}=0$ 
in the isospin limit and $|\vec q_c| = \omega_c$ at $Q^2 = 0$, we find that
\begin{equation}
W_{2,\alpha}(Q^2 \to 0) = \frac{1}{\vec q^2} \sum  |\bra{X} q\cdot A_{\alpha} \ket{N}|^2\,.
\label{eq:w2-2}
\end{equation}

According to Refs.~\cite{mf1,mf2,mf3},
the divergence equations for the axial currents give
\begin{equation}
|\bra{X(p')} q \cdot A^a \ket{N(p)}|^2 = 
f_\pi^2 m_\pi^4 |\bra{X(p')} \hat \pi^a \ket{N(p)}|^2 ,
\label{eq:pcac1}
\end{equation}
where $f_\pi$ ($m_\pi$) is the pion decay constant (pion mass), and 
$\hat\pi^a$ is the normalized interpolating pion field.
Furthermore, the matrix element $\bra{X(p')}\hat\pi^a\ket{N(p)}$ at $Q^2=0$ can be expressed as 
\begin{equation}
|\bra{X(p')} \hat \pi^a \ket{N(p)}|^2 = 
\frac{2\omega_c}{m_\pi^4}|{\cal T}_{\pi^a N \to X} (0)|^2 .
\label{eq:pcac2}
\end{equation}
Here, ${\cal T}_{\pi^a N \to X}(q^2)$ is the T-matrix element
of the $\pi^a(q) + N(p) \to X(p')$ reaction in the $\pi N$ center-of-mass frame 
(i.e., in the frame FM2), 
where the incoming pion can be off-mass-shell $q^2 \not= m_\pi^2$.
Using Eqs.~(\ref{eq:pcac1}) and~(\ref{eq:pcac2}), we have at $Q^2=0$,
\begin{eqnarray}
\frac{1}{\vec q^2} \sum
|\bra{X(p')} q \cdot A^a \ket{N(p)}|^2 
 &=& \frac{1}{\vec q^2} f_\pi^2 (2\omega_c)\sum  |{\cal T}_{\pi^a N \to X}(0)|^2
\nonumber\\
&\sim&
 \frac{1}{\vec q^2} f_\pi^2 (2\omega_c)\sum  |{\cal T}_{\pi^a N \to X}(m_\pi^2)|^2 
\nonumber\\
&=&
 \frac{1}{\vec q^2} f_\pi^2 (2\omega_c) \frac{1}{2\pi}\frac{p\cdot q}{E_N\omega_c} \frac{E_N}{m_N}
\sigma_{\pi^a N \to X}
\nonumber\\
&=&
 \frac{f_\pi^2}{\pi\omega} \sigma_{\pi^a N \to X},
\end{eqnarray}
where $\sigma_{\pi^a N \to X}$ is the total cross section of 
the on-shell $\pi^a + N \to X$ reactions, 
and we have used the relation
 ${\cal T}_{\pi^a N \to X}(q^2 = 0)\sim {\cal T}_{\pi^a N \to X}(q^2=m_\pi^2)$,
which is a consequence from the PCAC hypothesis~\cite{coleman},
and $\vec q^2 = \omega^2$.
From the fact that $\ket{\pi^\pm}=\mp(1/\sqrt{2})(\ket{\pi^1}\pm i\ket{\pi^2})$
and $\ket{\pi^0}=\ket{\pi^3}$, we finally have
\begin{equation}
W_{2,\alpha} = 
\left\{
\begin{array}{cl}
\displaystyle \frac{2f_\pi^2}{\pi\omega }\sigma_{\pi^{+} N \to X}  
&(\text{for }\alpha =\text{CC}\nu), \\
\\
\displaystyle \frac{2f_\pi^2}{\pi\omega }\sigma_{\pi^{-} N \to X}  
&(\text{for }\alpha =\text{CC}\bar\nu), \\
\\
\displaystyle \frac{ f_\pi^2}{\pi\omega }\sigma_{\pi^0 N \to X}  
&(\text{for }\alpha =\text{NC}\nu,~\text{NC}\bar\nu). \\
\end{array}
\right.
\label{eq:w2-pcac}
\end{equation}
The results so far obtained is essentially same as
in the papers by Adler~\cite{adler} and also by Paschos {\it et al.}~\cite{pas1,pas2,pas3}.
From Eqs.~(\ref{eq:crs2}) and~(\ref{eq:w2-pcac}), one can evaluate 
neutrino-induced forward meson production reactions at $\theta = 0$
using the $\pi N \to X$ total cross sections.

\section{Results and discussions}
\label{sec:result}

To evaluate Eq.~(\ref{eq:w2-pcac}), we need inputs of $\pi N$ reaction total cross sections.
In this work, we employ those obtained from the DCC approach 
developed by the authors~\cite{aip11,knls12}.
This approach is based on a DCC model~\cite{msl07},
within which the couplings among relevant meson-baryon reaction channels including the
three-body $\pi\pi N$ channel are fully taken into account, so that 
the scattering amplitudes satisfy the two-body as well as three-body unitarity.
The scattering amplitudes of $\pi N \to X$ with
$X = \pi N, \pi\pi N,\eta N, K\Lambda, K\Sigma$ are then
constructed through a global analysis of pion- and photon-induced
$\pi N$, $\eta N$, $K\Lambda$, $K\Sigma$ production reactions off the nucleons up to 
$W=2$ GeV.
Details of this analysis will be reported elsewhere~\cite{knls12}.

\begin{figure}[t]
\begin{center}
\includegraphics[width=\textwidth,clip]{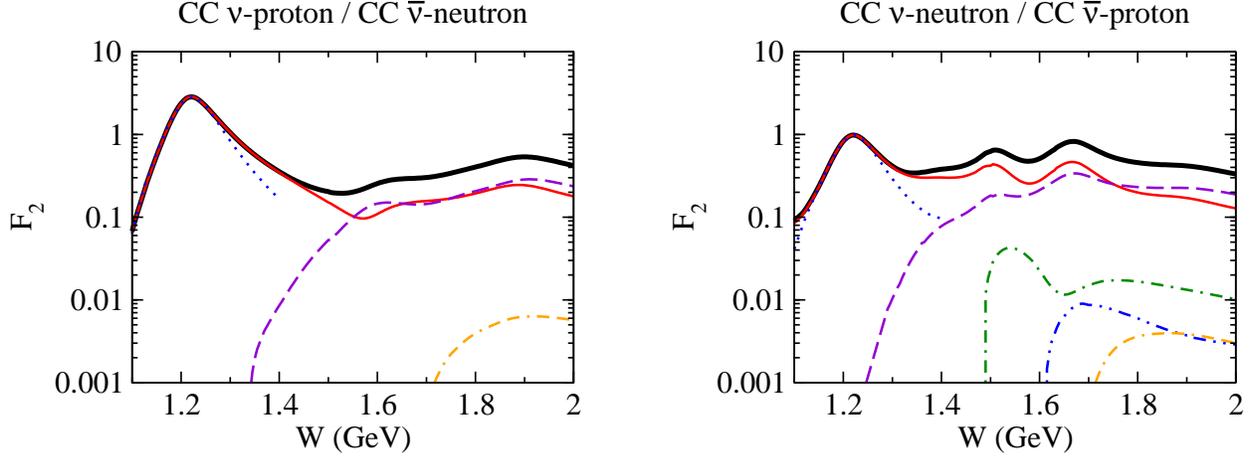}
\caption{(color online)
$W$-dependence of the $F_2$ function for CC neutrino-nucleon meson-production reactions 
at the limit $Q^2\to 0$, plotted for $W$ from the $1\pi$-production threshold up to 2 GeV.
The left (right) panel is for the CC $\nu$-proton or CC $\bar \nu$-neutron
(CC $\nu$-neutron or CC $\bar \nu$-proton) reactions.
Each curve is:
(thick solid curves) Total contribution from $X = \pi N, \pi \pi N, \eta N, K\Lambda, K\Sigma$ 
production reactions; 
(thin solid curves) contribution from $X = \pi N$ only;
(dashed curves) contribution from $X = \pi \pi N$ only;
(dashed-dotted curves) contribution from $X = \eta N$ only;
(dashed-two-dotted curves) contribution from $X = K \Lambda$ only;
(two-dashed-dotted curves) contribution from $X = K \Sigma$ only.
As a comparison, results from the SL model~\cite{sl3}, in which $F_2$ is directly calculated without relying on the PCAC hypothesis, are also shown as dotted curves.
}
\label{fig:cc}
\end{center}
\end{figure}

\begin{figure}[t]
\begin{center}
\includegraphics[width=0.5\textwidth,clip]{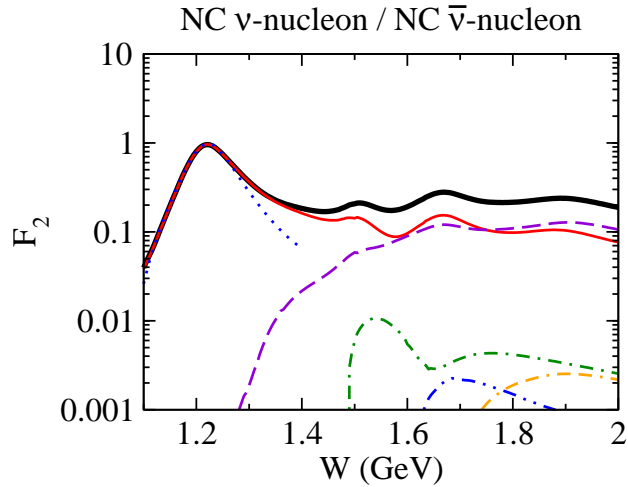}
\caption{(color online)
$W$-dependence of the $F_2$ function for NC neutrino-nucleon meson-production reactions 
at the limit $Q^2\to 0$, plotted for $W$ from the $1\pi$-production threshold up to 2 GeV.
The meaning of each curve is same as Fig.~\ref{fig:cc}.
}
\label{fig:nc}
\end{center}
\end{figure}

We present the structure functions $F_2$ of the neutrino-nucleon reactions 
(Fig.~\ref{fig:cc} for CC reactions and Fig.~\ref{fig:nc} for NC reactions).
Here $F_2$ is a dimensionless quantity defined by $F_2 = \omega W_2$.
It is found that the contribution of the $\pi N$ production reactions
dominate $F_2$ below $W=1.5$ GeV, while above that energy, the contribution of 
$\pi \pi N$ production reactions becomes comparable with $\pi N$, indicating the importance
of the $\pi \pi N$ reactions in the nucleon resonance region beyond $\Delta(1232)$.
On the other hand, contribution of $\eta N$, $K\Lambda$, and $K\Sigma$ reactions are
much smaller [$O(10^{-1})$-$O(10^{-2})$] than that of $\pi N$ and $\pi \pi N$,
which is similar to cross sections for meson production reactions 
with pion and photon beams.
In the same figures, we also present results from the Sato-Lee (SL) model~\cite{sl1,sl3,msl05} (dotted curves).
This model aims to describe $\pi N$ production reactions in the $\Delta(1232)$ region
and thus contains only $\Delta(1232)$ as resonance contributions.
It is noted that $F_2$ functions for the SL model shown in the figures
are {\it not} obtained via the PCAC hypothesis as discussed in
Sec.~\ref{sec:form-1}.
The SL model directly gives the $F_2$ functions because it
consists of both the vector and axial currents, and
reasonably reproduce available neutrino-induced pion production data~\cite{sl3}.
Comparing dotted curves with thin solid curves that are the full $\pi N$ 
production reactions up to $W = 2$ GeV, 
we clearly see that contributions from other than $\Delta(1232)$, 
e.g., higher resonances and/or backgrounds, become relevant above $W = 1.3$ GeV.
Also, the good agreement between the thin solid and dotted curves for
$W\lesssim 1.3$ GeV indicates a reliability of calculating $F_2$ 
from the $\pi N \to X$ total cross sections with the PCAC hypothesis.

It is noted that, even above the $\Delta(1232)$ region,
the $F_2$ function still has bump structures and is not a monotonous
function in $W$.
This non-monotonic behavior of $F_2$ comes from high-mass nucleon resonances, 
which are expected to exist up to $W \sim 2.5$ GeV.
An appropriate treatment of such behavior, as successfully done in our DCC approach, may be crucial 
for reducing systematic errors
in atmospheric and accelerator experiments to determine neutrino parameters.

\section{Summary and prospects for future directions}
\label{sec:summary}

As a first step toward developing a reaction model that enables a comprehensive 
description of neutrino-nucleon reactions in the nucleon resonance region, 
we have applied for the first time the DCC approach developed in Refs.~\cite{msl07,aip11}
to the neutrino-induced forward meson-production reactions off the nucleons,
$l + N \to l' + X$ with 
$(l,l') =(\nu_e,e^-), (\bar \nu_e,e^+), (\nu_e,\nu_e), (\bar \nu_e,\bar \nu_e)$ 
and $X = \pi N, \pi \pi N, \eta N, K\Lambda, K\Sigma$, in the energy region from the $\pi N$ 
threshold up to $W = 2 $ GeV.
This has been achieved by relating divergence of the axial-current matrix elements 
$\bra{X}\partial_\mu j_A^\mu \ket{N}$ at $Q^2 \to 0$ 
to the $\pi N \to X$ reaction amplitudes from the dynamical coupled-channels model 
through the PCAC hypothesis.

We have presented the $F_2$ structure functions for $l + N \to l' + X$ 
and investigated contributions of production reactions for each $X$.
It is found that above $W=1.5$ GeV, the contribution of $\pi \pi N$ production reactions 
becomes comparable with $\pi N$.
Also, our results suggest that a naive extrapolation of the DIS cross sections 
down to the nucleon resonance region, which is often performed in analyses of 
atmospheric and accelerator experiments, may be better to be replaced by more realistic 
reaction cross sections for precise determination of neutrino parameters.

The next step will be extending our dynamical coupled-channels model to directly analyze
neutrino reactions without relying on the PCAC hypothesis, 
so that we can investigate neutrino reactions at any finite $Q^2$.
This project is underway and will be reported elsewhere.

\begin{acknowledgements}
The authors thank Y.~Hayato, M.~Hirai, S.~Kumano, K.~Saito, and M.~Sakuda
for fruitful discussions at J-PARC Branch of KEK Theory Center.
HK acknowledges the support by the HPCI Strategic Program 
(Field 5 ``The Origin of Matter and the Universe'') of 
Ministry of Education, Culture, Sports, Science and Technology (MEXT) of Japan.
SXN is the Yukawa Fellow and his work is supported in part by Yukawa Memorial Foundation,
the Yukawa International Program for Quark-hadron Sciences (YIPQS),
and by Grants-in-Aid for the global COE program 
``The Next Generation of Physics, Spun from Universality and Emergence'' from MEXT.
TS is supported by JSPS KAKENHI (Grant Number 24540273).
This work is also supported by the U.S. Department of Energy, Office of Nuclear Physics Division, 
under Contract No. DE-AC02-06CH11357.
This work used resources of the National Energy Research Scientific Computing Center, 
which is supported by the Office of Science of the U.S. Department of Energy 
under Contract No. DE-AC02-05CH11231, and resources provided on ``Fusion,'' 
a 320-node computing cluster operated by the Laboratory Computing Resource Center 
at Argonne National Laboratory.
\end{acknowledgements}

\end{document}